\documentclass[twocolumn]{aastex7x}
\usepackage[utf8]{inputenc}
\usepackage{amsmath}

\def\kms{${\rm km\; s^{-1}}$}
\def\kmss{${\rm km\; s^{-2}}$}

\def\name{WD~J005311}
\def\namex{J0053}

\begin{document}

\title{The Highly Variable Wind from WD~J005311, the Stellar Remnant of the Peculiar Galactic Supernova of 1181}

\author[orcid=0009-0007-6386-151X]{C. Alexander Thomas}
\affiliation{Department of Physics and Astronomy, University of Notre Dame, Notre Dame, IN 46556}
\email{cthoma25@nd.edu}

\author[0000-0003-4069-2817]{Peter Garnavich} 
\affiliation{Department of Physics and Astronomy, University of Notre Dame, Notre Dame, IN 46556}
\email{pgarnavi@nd.edu}

\author[0000-0003-4773-4602]{Charlotte Wood}
\affiliation{Department of Physics and Astronomy, University of Notre Dame, Notre Dame, IN 46556}
\affiliation{Department of Physics, North Carolina Agricultural and Technical State University, Greensboro, NC 27411 }
\email{cwood@ncat.edu}

\author[0000-0003-1435-3053]{Richard Pogge}
\affiliation{Department of Astronomy, The Ohio State University, Columbus, OH 43210}
\affiliation{Center for Cosmology \&\ AstroParticle Physics, The Ohio State University, Columbus, OH 43210}
\email{pogge.1@osu.edu}

\author[0000-0003-4917-3685]{Lara Arielle Phillips}
\affiliation{Department of Physics and Astronomy, University of Notre Dame, Notre Dame, IN 46556}
\email{lphilli2@nd.edu}

\begin{abstract}

WD~J005311 is the peculiar stellar remnant of the Galactic supernova from 1181, and appears to have been the merger of two white dwarfs.  We present time-resolved spectroscopy of WD~J005311 showing emission line variability on a wide range of time-scales. The strong O~VI emission feature displays line profile variations (LPVs) on two distinct velocity scales. Broad variations with amplitudes of $\pm10$\%\ of the line flux are seen over the entire O~VI line. These broad LPVs likely arise from instabilities in the line-driven wind produced in many Wolf-Rayet stars. There is a hint of coherent structure in the broad LPVs that is consistent with rotation over roughly two hours, although the features survive for less than a full cycle.  Low-amplitude, narrow LPVs are also detected within the central $\pm5000$~\kms\ of the O~VI line. We associate these features with an unstable disk formed from the rigidly rotating magnetosphere (RRM) of the remnant white dwarf. We also analyze archival near-ultraviolet photometry of WD~J005311 and find a pseudo-periodic oscillation with an hour-long time-scale that maybe associated with the breakout instability of the RRM disk.

\end{abstract}

\keywords{\uat{Stellar mergers}{2157} --- \uat{Stellar winds}{1636} ---  \uat{Wolf-Rayet stars}{1806} --- \uat{Magnetic variable stars}{996} --- \uat{White dwarf stars}{1799} --- \uat{Supernova remnants}{1667}}

\section{Introduction}

White dwarfs (WDs) are degenerate remnants of low mass stars. The merger of WD binaries is expected to be common in the Galaxy \citep{branch95} and such mergers may be the major source of type~Ia supernova explosions \citep{iben84,webbink84}. Other, less explosive outcomes, are also possible depending on the masses and evolution of the binary system \citep[e.g.][]{dan12}.

\citet{gvaramadze19} identified an unusual star that may be a white dwarf binary merger remnant.  \name\footnote{\citet{kashiyama19} designated the star \name, although its white dwarf nature is not absolutely established. } (\namex\ hereafter) displays a spectacular spectrum of broad, high-ionization carbon and oxygen emission features. Its photospheric temperature is about 200000$^\circ$K and the lines suggest a wind velocity of 16000~\kms\ . 

Infrared emission around \namex\ was first suspected to be a planetary nebula (PN) by Dana Patchick (Deep Sky Hunters Collaboration) and was given the name Pa30 in the compilation of new PN by \citet{kronberger16}. 
\citet{oskinova20} noted that the abundances derived from the X-ray emission of \namex\ and its surroundings are consistent with a WD merger embedded in a supernova remnant. A remarkable [S~II] emission nebulosity around the star was mapped by \citet{fesen23}. \citet{ritter21}, and later \citet{schaefer23} associated the remnant with the Supernova of 1181, and the explosion appears consistent with a type~Iax supernova (SNIax) \citep{jha17,ritter21}.

The observed properties of \namex\ are quite different from other suspected post-merger WDs \citep[e.g. WD~0525+526;][]{sahu25}. The unique characteristics of \namex\ suggests that it is still evolving after the merger and partial deflagration as a SNIax. Models of WD mergers \citep{shen12,schwab12,schwab16,schwab21} show that they may undergo thousands of years of viscous and thermal evolution post-merger. Eventually, a SNIax deflagration may be triggered that ejects a fraction of the merged remnant. Specific models for Pa30 imply that some of the material remained bound and fell back on to the stellar remnant \citep{ko24}. The deflagration and fallback heats the outer layers that leads to a hot envelope surrounding the remnant WD \citep{piro26}. It is possible that carbon burning ignites at the base of the envelope and provide a long-term heating source \citep{piro26,ko26}. 

Support for the delay between the merger and SNIax explosion comes from the discovery of a blue-shifted Na~I absorption feature in the spectrum of \namex . This may be material ejected during the merger, but has not yet been swept up by the faster moving supernova ejecta \citep{garnavich26}.

The post-merger object is likely to be rapidly rotating and possibly possess a strong magnetic field  \citep{gvaramadze19}. Spectra of \namex\ are similar to that of a hot, oxygen-rich Wolf-Rayet (W-R) star, but with an unprecedented wind velocity. Its strongest emission line at optical wavelengths is the O~VI doublet at 381.1/383.4~nm. Time-resolved spectra showed significant variability in the emission substructure of the line over a 30-minute time-scale \citep{garnavich20}. Variable line substructure is often seen in W-R stars, but over longer time-scales and with smaller amplitudes than observed in \namex\  \citep[e.g.][]{lepine99a}.

 \citet{lykou23} made detailed models of the optical and ultraviolet spectra of \namex\ and found a low Ne/O abundance implying that the merger did not involve a Ne-O WD. They also confirmed the emission line variability noted by \citep{garnavich20}. A slow dimming of \namex\ over the past 100 years was discovered by \citet{lykou23} and \citet{schaefer23} using a photographic plate archive. Fading over the past 20 years was also detected in the mid-infrared bands \citep{lykou23}.

\begin{figure}[t]
    \centering
    \includegraphics[width=\columnwidth]{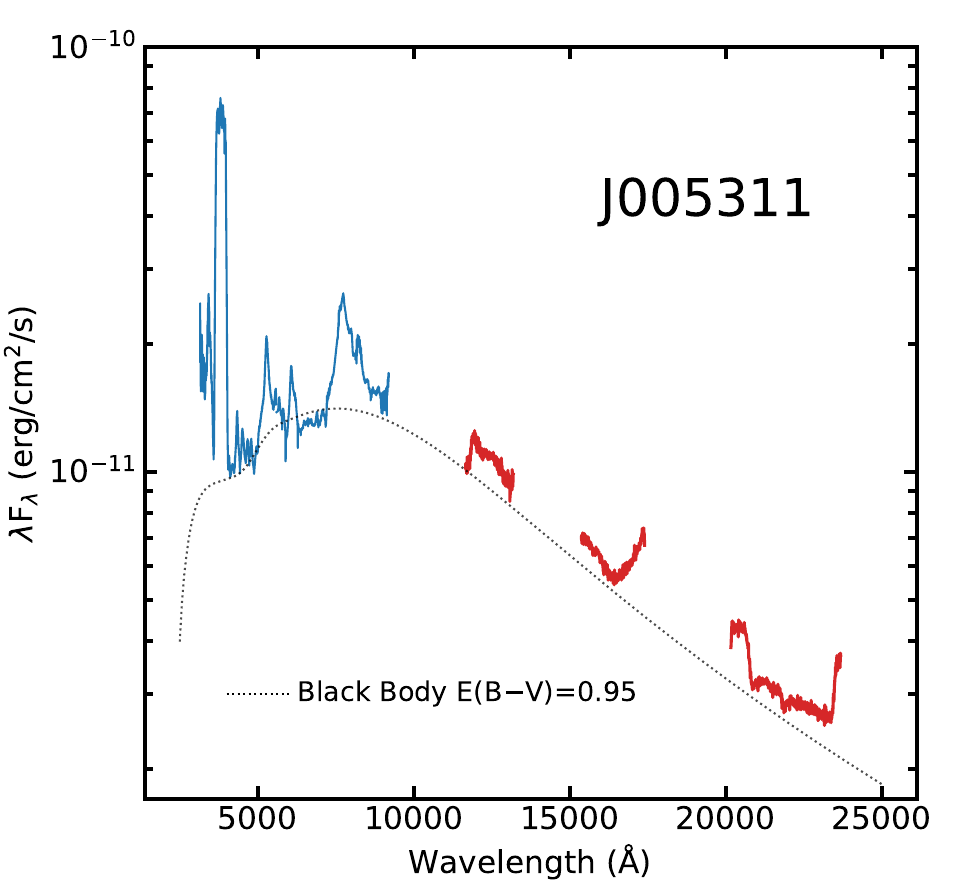}\hfill
    \caption{Spectra of \namex\ from 3200~\AA\  to near-infrared wavelengths. The IR spectra were taken in the $J$, $H$, and $K$ atmospheric windows and displayed as red lines. Weak emission features are detected in the near-IR, but are cutoff in these narrow bands due to their high Doppler widths. The average LBT/MODS spectrum from 10/16/2022 is shown as a blue line. The dotted line shows a hot black body with dust extinction of $E(B-V)=0.95$~mag.  }
    \label{irspec}
\end{figure}

The expected rapid rotation and strong magnetic field likely results in a rigidly rotating magnetosphere \citep[RRM;][]{townsend05,ud-doula08} where the wind is channeled along the rotating magnetic field out to the Alf\'ven radius. If the stellar rotation is sufficiently high so that the Kepler corotation radius is inside the Alf\'ven radius, a centrifugal magnetosphere \citep{petit13} results. In this case, the wind is channeled to an unstable disk which breaks out to drive an equatorial outflow \citep{ud-doula08}. This unstable disk and outflow might be detectable in \namex . At high latitudes beyond the Alf\'ven radius, the field lines are open and the wind may continue to accelerate through multi-line scattering as seen in high mass W-R stars \citep[e.g. CAK;][]{cak75}. 

Here, we present several epochs of high-cadence time-series optical spectroscopy of \namex , and analyze their characteristics. We also study the first near infrared (IR) spectra of \namex\ and analyze archival ultraviolet (UV) and optical data.

\section{Data}

\subsection{Optical Spectra}

\begin{deluxetable*}{lcccccc}
\centering
\tablewidth{0pt}
\tablecaption{J005311 Optical Spectroscopy \label{Observations}}
\tablehead{
\colhead{Date (UT)} & \colhead{MJD} & \colhead{Exposure Time} & \colhead{Number of} & \colhead{Telescope/}  & \colhead{Wavelength Range} \vspace{-0.2cm}
\\
\colhead{(mm/dd/yyyy) } & \colhead{(days)} & \colhead{(seconds)}& \colhead{Spectra} & \colhead{Instrument}  & \colhead{(\AA )}
}
\label{table}
\startdata
09/11/2020 & 59103 & 540.0 & 5  & LBT/MODS & 3150-5650  \\
10/10/2020 & 59132 & 152.6 & 30 & LBT/MODS & 3150-5670  \\
10/12/2020 & 59134 & 152.6 & 23 & LBT/MODS & 3150-5650  \\
11/21/2020 &  59174 & 152.6 & 30 & LBT/MODS & 3150-5650  \\
01/04/2021 & 59218 & 152.6 & 33 & LBT/MODS & 3150-6200  \\
01/06/2021 & 59220 & 152.6 & 34 & LBT/MODS & 3150-6200  \\
09/05/2021 & 59462 & 180.0 & 54 & KECK/LRIS  & 3163-5599  \\
09/09/2021 & 59466 & 152.6 & 37 & LBT/MODS & 3150-5590  \\
10/16/2022 & 59868 & 150.0 & 42 & LBT/MODS & 3150-5590  \\
\enddata
\end{deluxetable*}

We obtained spectra of \namex\ using the Large Binocular Telescope (LBT) and the twin Multi-Object Dual Spectrographs \citep[MODS;][]{pogge12} on a total of 8 dates listed in Table~\ref{table}. The observations from September 2020 are already discussed in \citep{garnavich20}. The LBT/MODS2 time-series consisted of 20 to 40 exposures of 153~s in length with typical overheads of 63~s. Thus, the cadence of the LBT spectral time-series was 220~s, and the time-series typically spanned two hours. The LBT spectra were taken in dual grating mode using a 0.8~arcsec wide slit providing a spectral resolving power of approximately 2000, or a full-width at half-maximum (FWHM) of 200~\kms\ at 3900~\AA .  For the blue spectra obtained with the LBT, the dispersion was resampled to 0.5~\AA\ per pixel. 

In addition, we obtained a night of observations using the Keck Observatory and the Low Resolution Imaging Spectrometer (LRIS), which took place on 2021 September 5 (UT). The blue spectrograph employed the 400/3400 grism and a 0.7~arcsec wide slit resulting in a resolving power of 1000 or, equivalently, a feature width of 300~\kms\ (FWHM). The seeing over the time series varied between 0.7 and 0.9 arcsec (FWHM). The spectra were resampled to a dispersion of 1.0~\AA\ per pixel. The 54 images using 180s exposures spanned a total of 3.5~hours with short gaps due to instrument issues.

Using reduction packages in PyRAF/IRAF\footnote{Interactive Reduction and Analysis Facility \citep[IRAF;][]{tody86,tody93} was distributed by the National Optical Astronomy Observatories, which are operated by the Association of Universities for Research in Astronomy, Inc., under cooperative agreement with the National Science Foundation. PyRAF \citep{pyraf12} is a command language for running IRAF tasks in a Python like environment.}, the CCD spectral images were extracted into one-dimensional spectra. The MODS spectrograph is split between a blue and a red channel; since the wavelengths of interest were completely covered by the blue channel, we specifically focused on those images for data reduction. The LBT images were wavelength calibrated using an argon emission line lamp and were flux calibrated using the following spectrophotmetic standard stars: G191b2b for the January 4th, January 6th, September 9th and September 11th datasets, BD 28d4211 for the October 10th and October 12th datasets, Feige 100 for the November 21st dataset and Feige 25 for the Keck dataset. The MODS2 spectra had a higher signal-to-noise ratio (S/N) than the MODS1 images, so we concentrate our analysis on the MODS2 spectra.

\subsection{Near-Infrared Spectra}

We obtained near-IR spectra in the $J$, $H$, and $K$ bands on 2021 January 4 (UT), using the LUCI spectrographs on the LBT. The two LUCI spectrographs employed the N1.8 camera and G210 gratings with the 0.75~arcsec wide slit. \namex\ was observed at three separate grating tilts corresponding to the $J$-band (4th order), $H$-band (3rd order), and $K$-band (2nd-order) using the corresponding filters to separate the orders. 

The star was nodded at two positions along the slit in a ABBAABBA sequence with 60~s exposures taken for $J$ and $H$ and 30~s exposures obtained at the $K$ setting. Pairs of consecutive images were subtracted and the residuals extracted using PyRAF's TWODSPEC tools. Wavelength calibration was performed using the sky emission lines. Spectra of the nearby telluric standard star HIP6002 were obtained and used to remove telluric absorption features and flux calibrate the \namex\ spectra.

\subsection{HST Ultraviolet Observations}

\namex\ was observed with the Hubble Space Telescope (HST) using the Space Telescope Imaging Spectrograph (STIS; GO15864 P.I. Graefener). Spectra were taken with the G230L and G170L gratings providing coverage from 1150 to 3100~\AA\ using the TIMETAG mode of the FUV and NUV MAMA detectors. A comparison between these spectra and a model of \namex\ was published in \citet{lykou23}, and they identified two strong, broad emission lines in the ultraviolet (UV) as O~V 1371~\AA\ and 2788~\AA. The STIS spectra were obtained on two visits, 2020 Nov 4 and Nov 6, allowing study of line variability over 48 hours. Also, the TIMETAG mode allowed a search for rapid photometric variability in the UV.

We downloaded the reduced and extracted x1d spectra and tag files from the Milkulski Archives for Space Telescopes (MAST). To reduce the noise in the spectra, we averaged the individual exposures for each grating, resulting in a single FUV and a single NUV spectrum for each visit.

\begin{figure}[b]
    \centering
    \includegraphics[width=\columnwidth]{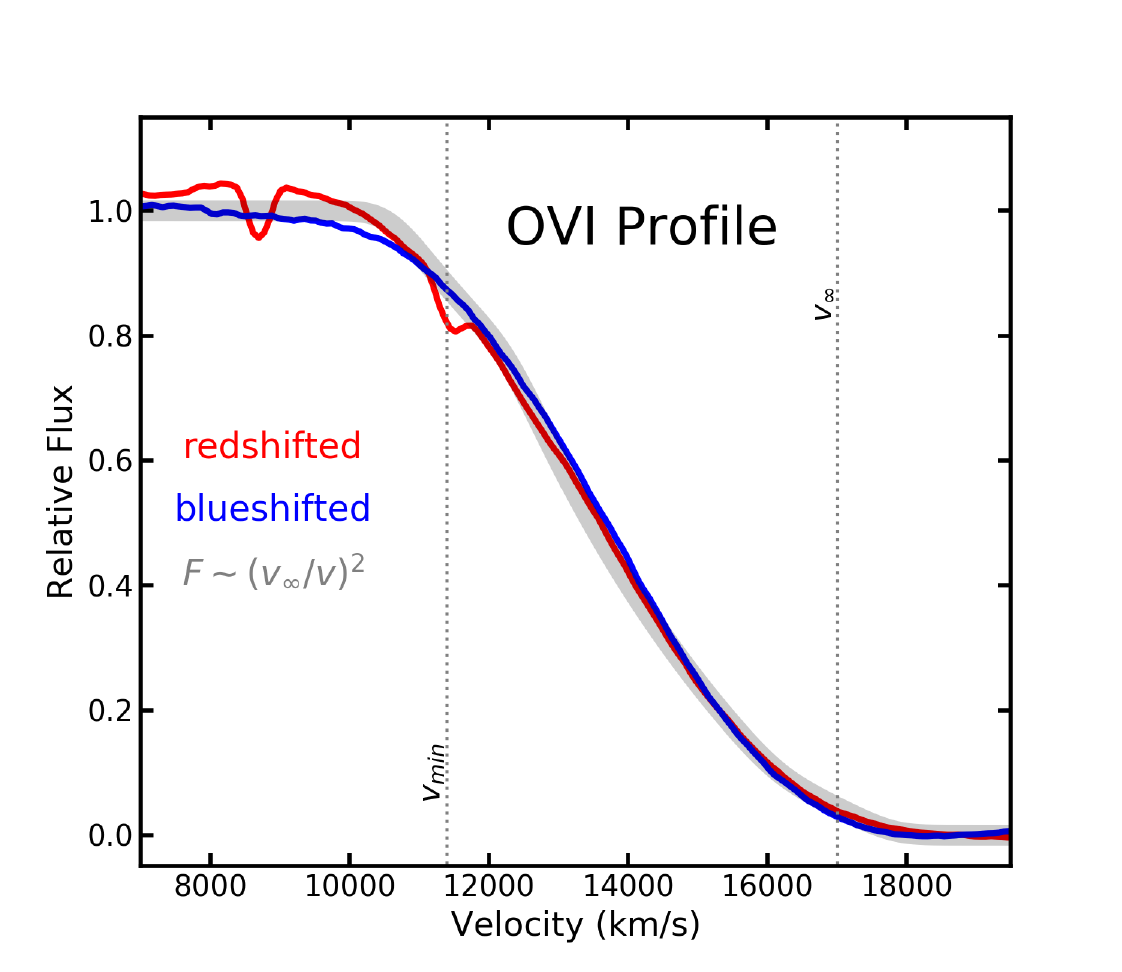}\hfill
    \caption{The O~VI emission line profile from the average of the individual Keck spectra. The velocity is relative to the assumed rest wavelength of 3822~\AA . The blue line shows the blue-shifted side of the line and the red is the shows the red-shifted side (note the two narrow interstellar Ca~II absorption lines). The gray band shows the line profile from a simple wind model where the flux falls off as $(v_\infty /v)^2$. $v_\infty$ is the wind velocity at large distances and $v_{min}$ is the velocity where O~VI emission first appears. To match the observations, the model has been smoothed using a pair of Gaussian functions separated by the difference in the O~VI doublet wavelengths.   }
    \label{profile}
\end{figure}

The tag files were read as FITS tables providing the time of the photon detection and its position on the detector. Using all the detections, we identified the stellar trace along the dispersion direction and extracted their detection times. We identified regions on each side of the stellar trace to estimate and subtract the background count rate. The count rate from \namex\ in the FUV was typically 17~cnts~s$^{-1}$, and in the NUV the rate averaged 93~cnts~s$^{-1}$. 

\section{Analysis}

The strongest emission line in the optical spectra of \namex\ is the extremely broad O~VI feature centered at 3822~\AA. This line is actually a doublet with components at 3811.4~\AA\ and 3834.2~\AA . The components are separated by 1790~\kms\ while the line is Doppler broadened by 32000~\kms , so the separation of the doublet is generally insignificant. For this analysis we will consider the rest wavelength of the feature to be 3822~\AA .  Identification of other emission features at optical and ultraviolet (UV) wavelengths can be found in \citet{gvaramadze19}, and \citet{lykou23}.

The LBT near-IR spectra are shown in Figure~\ref{irspec}, along with an LBT optical spectrum. The $H$-band spectrum reveals a rising flux starting at 1.65~$\mu$m that may be a blend of weak emission lines. Notably there are sharp edges of emission at both ends of the $K$-band. It is difficult to identify these lines as their central wavelengths are uncertain. If we assume a shape and width similar to the strong O~VI line in the UV, then their wavelengths are approximately 1.99~$\mu$m and 2.45~$\mu$m. Extensive modeling of the lines by \citet{lykou23} in the UV and optical show that most of the emission is identified as permitted transitions of high ionization oxygen, carbon, and neon. So we suspect that these two features in the $K$-band are dominated by Ne~VIII at 1.98~$\mu$m and 2.39/2.42~$\mu$m, based on the NIST atomic spectra database \citep{kramida24}. 

When combined with an LBT optical spectrum, the overall continuum is consistent with a hot black body and a color excess of $E(B-V)=0.95$,  which is consistent with the estimate by \citet{lykou23}.

\begin{figure}
    \centering
    \includegraphics[width=\columnwidth]{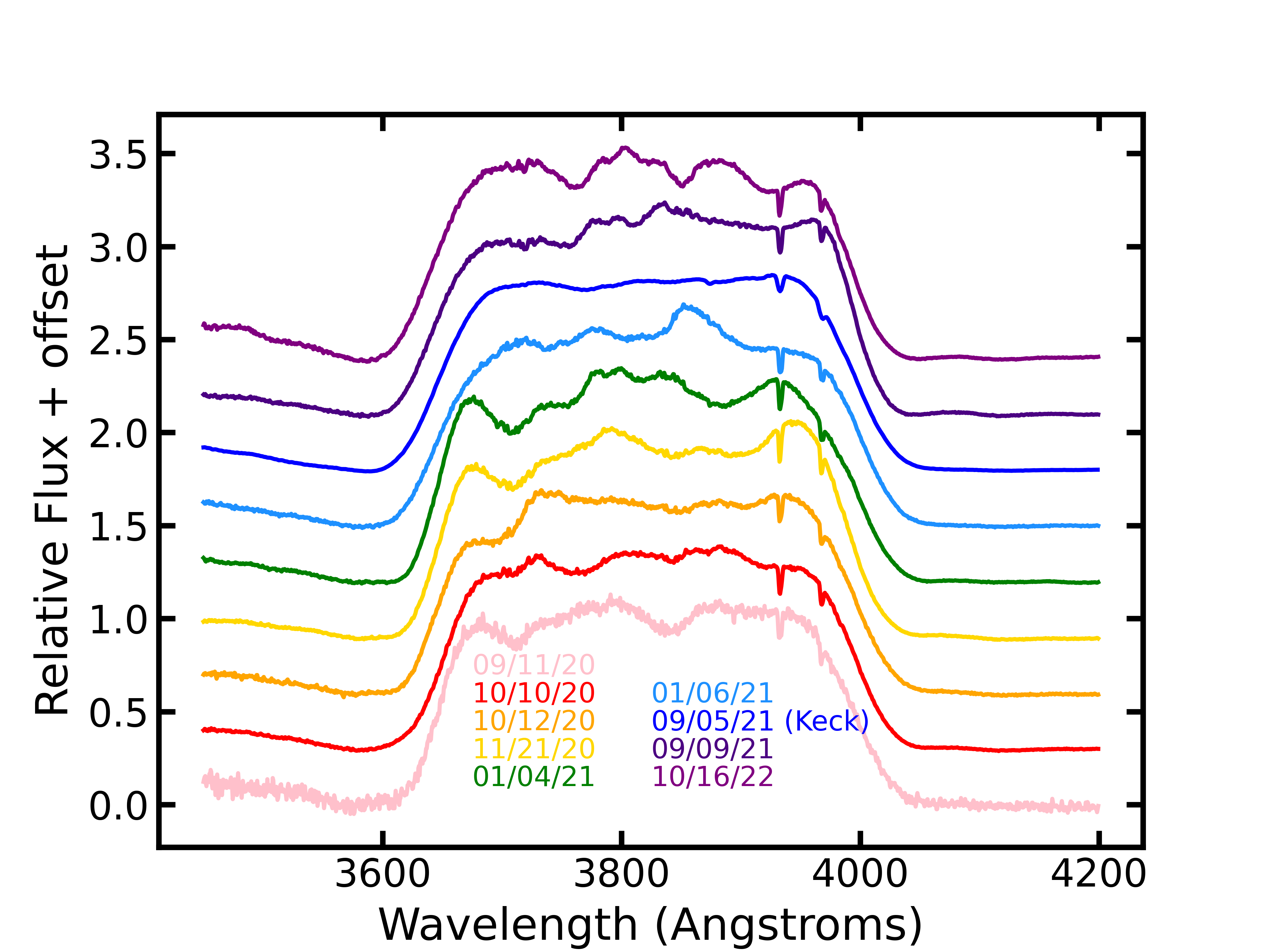}\hfill
    \caption{The average line profile of O~VI feature for each night. For clarity, the spectra have been offset vertically. Narrow interstellar calcium ``H+K'' absorption lines are seen on the red side of the broad emission. Variations in line shape at the $\pm 10$\%\ level are seen on time-scales of days to months, but no periodic pattern is seen.}
    \label{plotall}
\end{figure}

\subsection{Line Profile} \label{sec:line_shape}

Flat-top line profiles are often seen in W-R stars \citep[e.g. C III;][]{lepine99a} and these have been explained as shells of emission originating from an ion in a cooling and accelerating wind \citep{dessart05}.  At small radii the temperature exceeds the ionization energy of the ion of interest and it is only when the wind cools that the emitting ion becomes abundant.  The wind velocity, $v_{min}$, where this temperature is reached can be approximated as a thin emitting shell. The wind accelerates and the emission decreases producing fading wings.  Eventually the wind reaches an asymptotic velocity defined as $v_\infty$, and this is where the line emission reaches zero flux. Wind models are often parameterized as a function of radius $r$, using
\begin{equation}
v(r)=v_\infty (1-R_*/r)^\beta \;\; ,
\end{equation}
where $R_*$ is the stellar surface.  For $\beta=1$, the line flux is expected to fall off  as $(v/v_\infty)^{-2}$ between $v_{min}$ and $v_\infty$ \citep{owocki13}.  \citet{dessart05} have shown that the decay rate of the line wings is rather insensitive to the choice of $\beta$ for optically thin lines. 

\begin{figure}
    \centering
    \includegraphics[width=7cm]{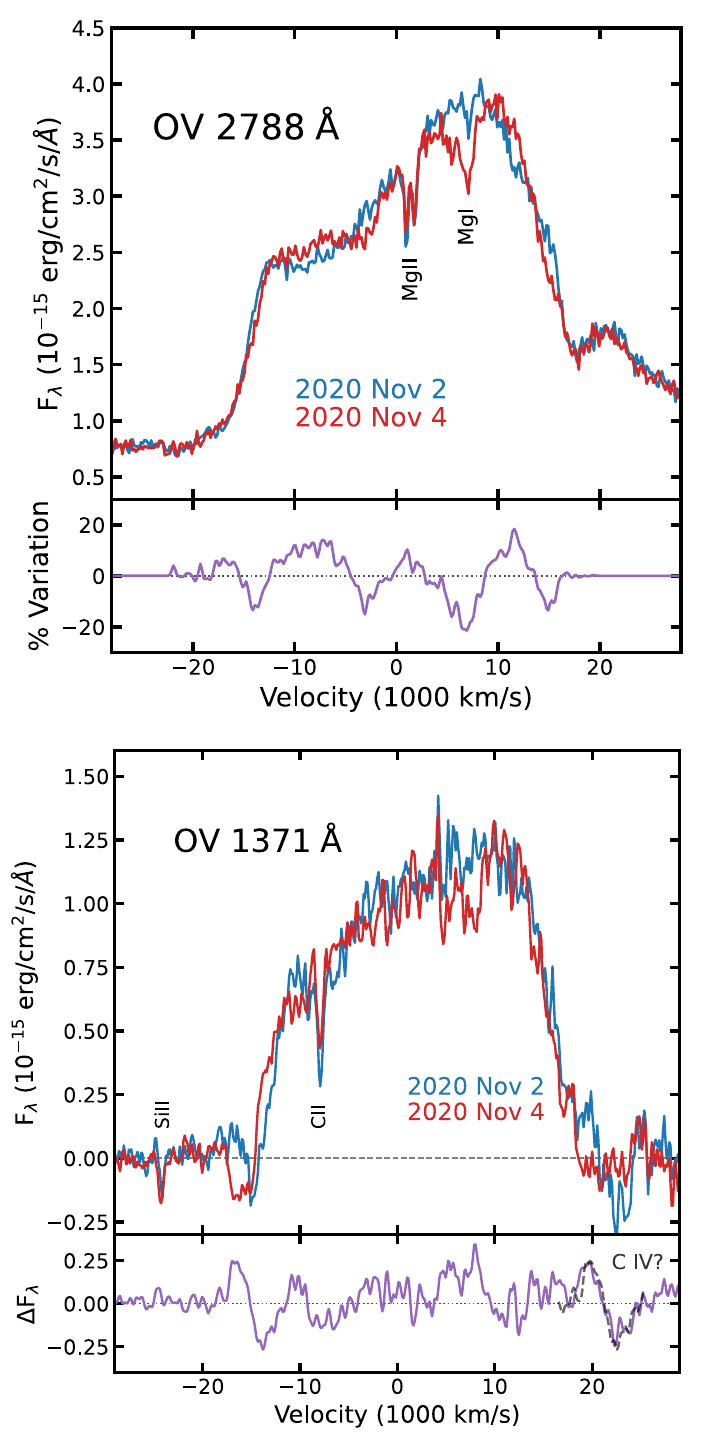}\hfill
    \caption{{\bf Top:} The NUV O~V emission line from the HST observations. The first visit is shown in blue and the second, two days later, is displayed in red. Narrow interstellar absorption lines of Mg~II and Mg~I are present. Reddening from dust combined with additional emission from Ne~VIII results in the emission appearing to rise toward the red. The lower panel shows the percent difference in the line flux between the two visits after removing the continuum contribution. {\bf Bottom:} The FUV O~V emission line displaying the profile variation over two days. A variable P-Cygni absorption is evident at the highest blue-shifted velocities. A similar profile change is seen at $+22000$~\kms\ that is likely from the blue-shifted C~IV 1548~\AA\ absorption as predicted by \citet{lykou23} .
    \label{hst} }
\end{figure}

From Figure~\ref{profile} we see that the observed line profile is indeed flat-topped and symmetric about the O~VI rest wavelength of 3822~\AA . This symmetry implies that the emission is optically thin.  The flux level remains nearly constant out to $v\approx 11000$~\kms\ corresponding  $v_{min}$ and the line reaches zero flux at about 18000~\kms . The intrinsic line shape is modified by the spectroscopic resolution, and by the fact that the emission line is a close doublet. Convolving the \citet{owocki13} model with a pair of Gaussian functions separated by 1800~\kms\ is needed for proper comparison with the data. Using this technique, we find an excellent match between the model and the observed line profile when $v_{min}=11400\pm 200$~\kms\ and $v_\infty = 17000\pm 200$~\kms . 

For optically thin lines, the velocity centroid of the emission, $v_e$, can be estimated by noting where the emission has reached 40\%\ of the peak flux \citep{dessart05}.  The width of the emitting region and turbulence in the wind can impact this estimate. In our data the flux has decayed to 0.40 of the peak at 14150$\pm 100$~\kms , meaning that $v_{e}/v_\infty \approx 0.8$. Thus, the O~VI emission in \namex\ is formed in the outer wind as is true of emission features in many W-R stars \citep{dessart05}. The width of the emitting zone is also likely narrow with $\Delta v_e$ between 0.1 and 0.2.

Similar results are obtained by analyzing the O~V 2788~\AA\ emission profile in the NUV. The redshifted side of the line is contaminated by Ne~VIII \citep{lykou23}, but the blueshifted edge provides a clean estimate of the wind parameters. We first corrected for the steep wavelength dependent extinction that distorts the line shape. $v_\infty = 17000\pm 300$~\kms\ and $v_{min}=12900\pm 300$~\kms . The $v_{min}$ value for O~V is 1500~\kms\ larger than for the O~VI emission. This is expected for the accelerating and cooling wind.

Solving for the radii of the shells from Equation~1, and $\beta=1$, the inner radius of the O~VI shell is 3.0~$R_\star$, with the bulk of the line emitting region at 6.0~$R_\star$. Similarly the inner radius of the O~V shell is 4.1~$R_\star$ for an assumed $\beta=1$.

\subsection{Long-Term Line Variations}

For each spectrum, we continuum subtracted and normalized the O~VI emission line. We then averaged the spectra from each night to study the shape variations on a time scale of weeks to months. The average spectra for each night are displayed in Figure~\ref{plotall}. Overall, the width and shape of the line remains consistent over the two years of monitoring. There are significant line profile variations (LPVs) at the $\pm 10$\%\ level between the observations. Even visits within a few days of each other show variations in positions and strengths of the bumps and dips on the top of the broad line. 

The amplitude of fluctuations on top of broad emission lines in in W-R stars tends to be of order 1\%\ of the peak flux \citep{lepine99a}. Variations with amplitudes as large as those in \namex\ have been seen when the W-R star is a component in a binary system. For example, WR~42 and WR~79 have O-type companions and orbital periods of about a week. \citet{hill00} has modeled the changes in the C~III emission lines in these stars as a combination of a typical smooth broad line with periodically shifting emission resulting from the collision of the winds from the two stars. There are no clear periodic variations in the O~VI line, however, the irregular sampling of the spectra makes detection of a long-term periodicity difficult.

\subsection{Line Variations in the UV}

LPVs are seen in the NUV and FUV O~V emission lines as well as the optical O~VI line. Figure~\ref{hst} displays the O~V 2788~\AA\ line on visits separated by two days. The profile fluctuated over velocity scales of about 5000~\kms\ and with amplitudes of $\pm$15\% .

The FUV O~V 1371~\AA\ emission displays a variable P-Cygni absorption at the far blue-shifted edge of the line (lower panel of Figure~\ref{hst}). The absorption falls below the continuum level for velocities blue-ward of $-14000$~\kms .  On the second visit the absorption is seen to expand to higher velocities, reaching to $-17200$~\kms . This matches the wind $v_\infty$ velocity measured from the O~VI and NUV O~V lines. A similar variation in absorption is seen at the extreme red side of the O~V 1371~\AA\ plot. At $+22000$~\kms, the varying absorption is greater than $v_\infty$ and unlikely to be associated with O~V, but it is probably from the blue edge of C~IV 1548~\AA . Spectral modeling by \citet{lykou23} (their figure~3) predicted deep P-Cygni absorptions in the FUV O~V and C~IV features.

\begin{figure}
    \centering
    \includegraphics[scale=0.35]{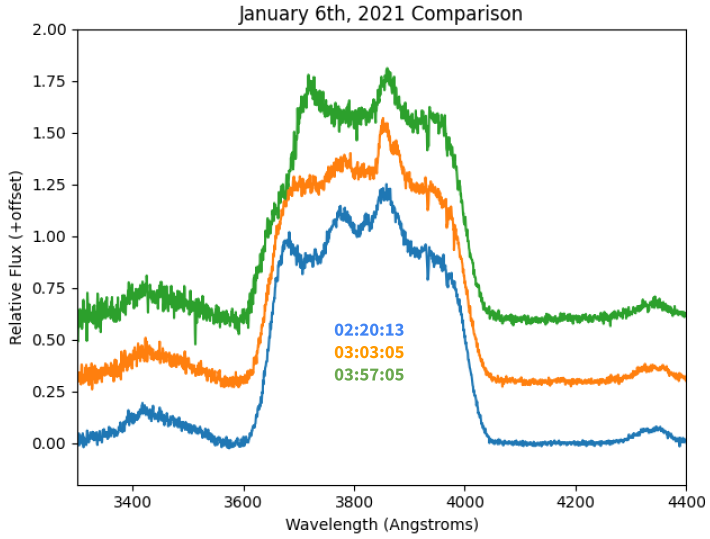}\hfill
    \caption{Three individual spectra of the main O~VI feature over the night of 2021 January 6. There is an offset between each individual spectrum and, as the spectrum gets higher, it gets later in the night. There are significant changes in the feature over 90-minutes as seen by the disappearance of a peak at the blue edge the broad emission, followed by the appearance of a peak slightly to the red of the original peak.}
    \label{example}
\end{figure}

\begin{figure*}
    \centering
    \includegraphics[scale=1.0]{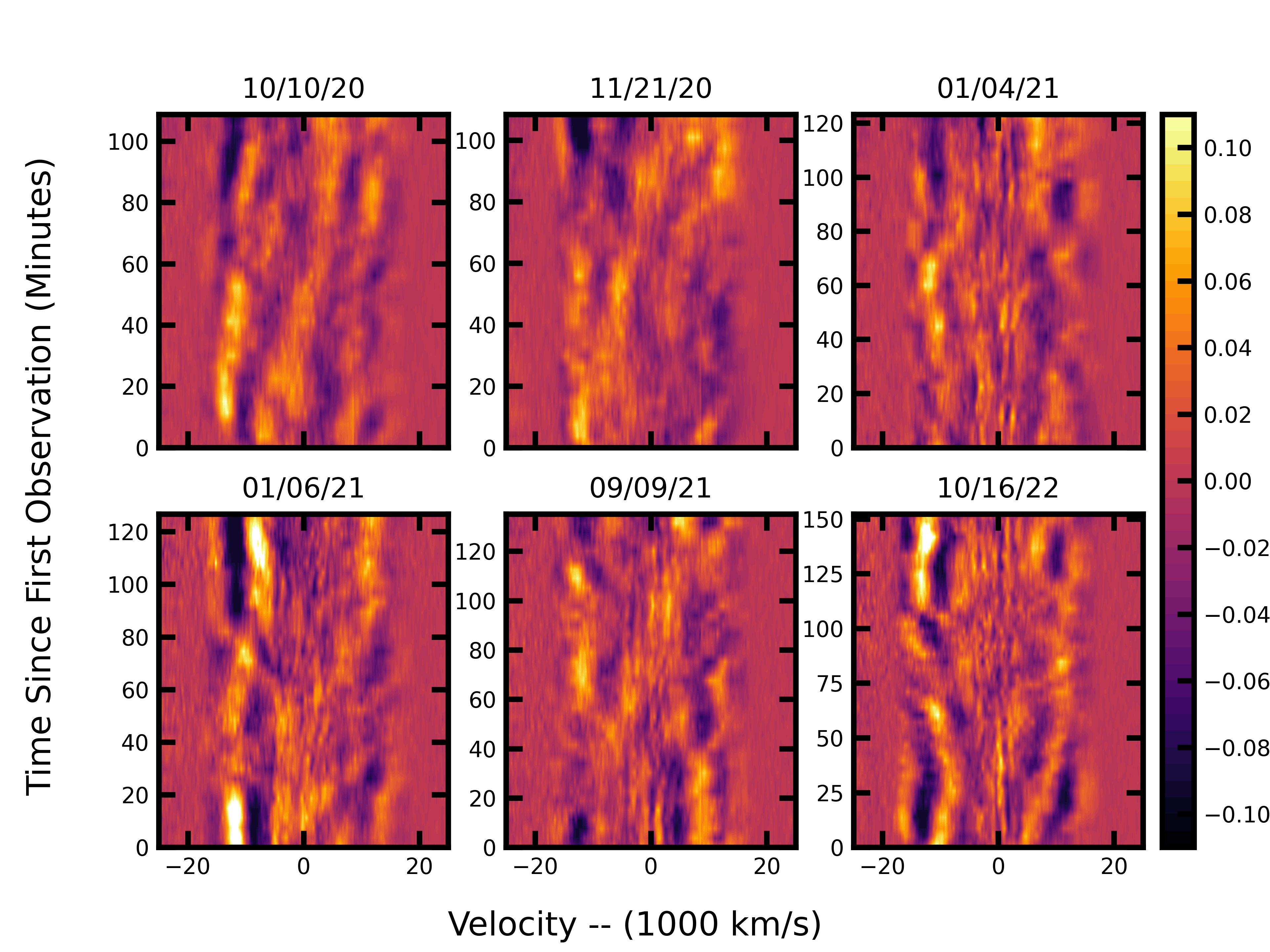}
    \caption{The residual time series spectra of \namex\ for each long time-series from the LBT. Along the x-axis is the velocity in 1000 km/s from the central wavelength of the O~VI feature (3822 {\AA}), the y-axis shows time since the first observation in minutes, and the color bar represents the relative intensity of the deviations from the average spectra for that night.}
    \label{residuals}
\end{figure*}

\subsection{Residual Time-Series}

\citet{garnavich20} noted a rapid variability in the O~VI emission feature during 2020 September observations. To characterize and quantify the variability, we examine several spectral time-series, each lasting approximately two hours with a typical cadence of 3~minutes. An example of the profile changes is shown in Figure~\ref{example}, where three representative O~VI line profiles covering 1.5~hours are compared. 

To display these variations, we calculate the residuals from the mean profile \citep{lepine99b}. This is achieved by estimating the continuum on each side of the emission line and subtracting a best fit linear function between the continuum regions. Each spectrum is then divided by its average flux measured between velocities of $\pm 12000$~\kms\ relative to its rest wavelength. The instantaneous residuals are calculated by subtracting each normalized spectrum from the mean spectrum for that night. The resulting residual time-series for each of the LBT data sets is shown in Figure~\ref{residuals}.

\begin{figure*}
    \centering
    \includegraphics[scale=1.0]{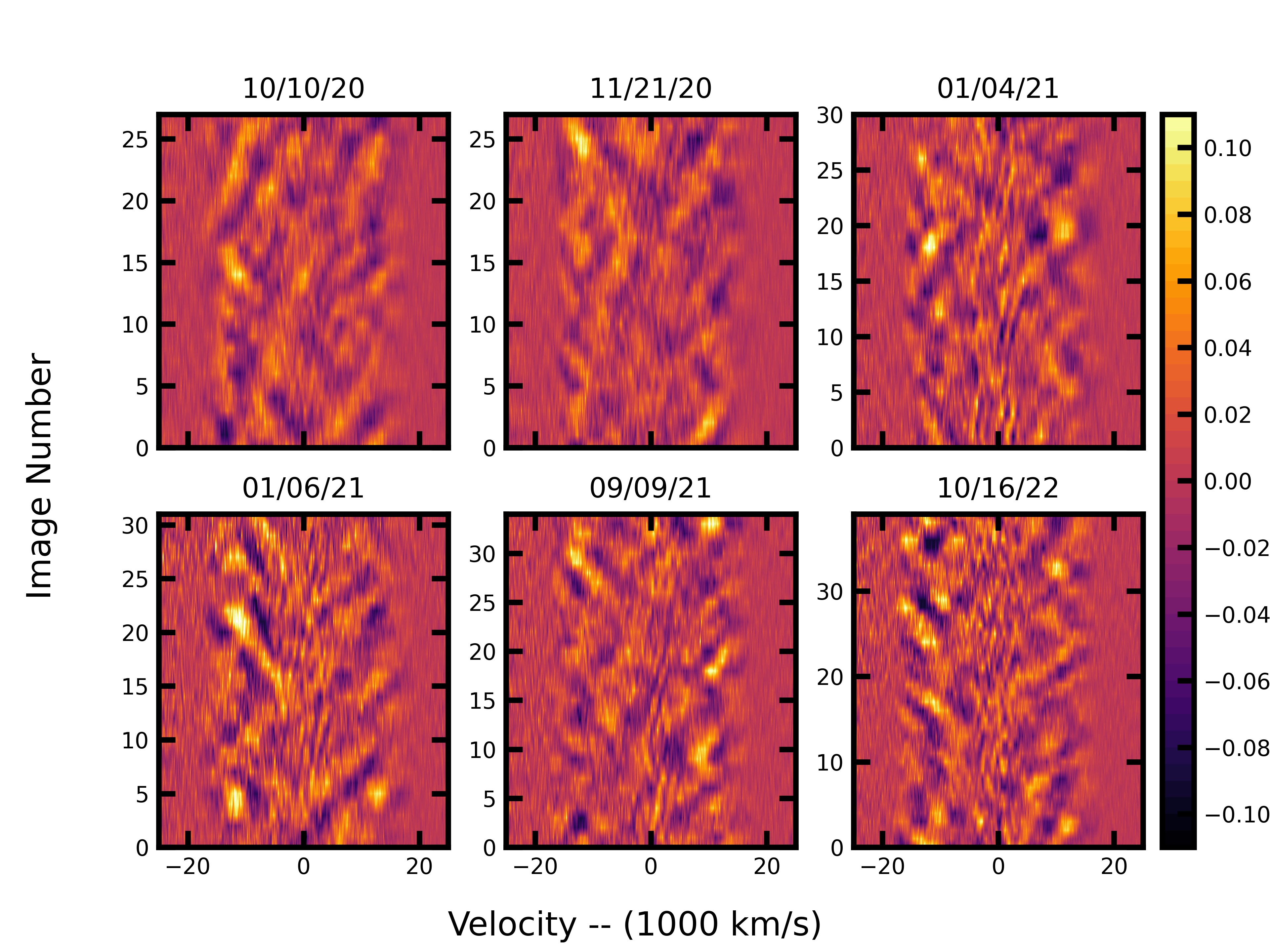}
    \caption{Instantaneous spectral variation gradient time series of J0053 for each night of the LBT data. In contrast to Figure~\ref{residuals}, the instantaneous gradient is constructed by differencing spectra separated by a fixed interval, as described in \citep{lepine99a}. Here, the time interval is 440~s. The x-axis shows the velocity relative to the O~VI rest wavelength of 3822{\AA} and the color bar represents the relative intensity of the deviations from the subtracted spectra. In this case, the y-axis denotes the image number of the time-series.}
    \label{instant}
\end{figure*}

Broad LPVs with amplitudes as large as 10\%\ of the peak line flux are detected. The broad fluctuations have widths of $\approx 6000$~\kms\ and are mainly seen at the highest red and blue shifted velocities.  Broad features are seen near the center of the O~VI emission, but they tend to vary on a slower time-scale than the features at the blue and red wings of the line.

The residual time series reveals presence of a second varying component made up narrow features that are not easily seen directly in the spectra.  Narrow features are primarily clustered toward the center of the O~VI line and consistently appear as two components. We suspect that the doubling is a result of the doublet nature of the O~VI emission. As the O~VI components are separated by $1800$~\kms , resolving the pair implies widths for the individual features of less than $\approx 900$~\kms . 

In general, neither the broad nor the narrow fluctuations appear to be consistently accelerating over the length of the observations. Accelerating features are often seen in spectra of W-R stars \citep[e.g.][]{lepine99a} and \citet{garnavich20} suspected an acceleration to be present in \namex . However, these LBT residual time-series shows no steady increase in velocity for the features over a two-hour span.

The narrow features are difficult to detect in the 2020 October nights, possibly due to unusually poor seeing conditions on those dates. But these surprising features are obvious in all the later time-series.

\subsection{Instantaneous Gradient Time-Series}

Interpretation of residual time-series images can be misleading. As noted by \citet{lepine99b}, the calculation of the residual time series can result in removal of slowly varying features. Subtracting the average forces single transients regions to be paired with a corresponding feature with an opposite sign so that the average flux is constant.


To avoid spurious correlations that result from applying the residual time-series technique, we also calculate the instantaneous gradient time series for each night as shown in Figure~\ref{instant}. As described in \citet{lepine99b}, the gradient is calculated from the difference between spectra separated by a fixed time $\Delta t$. For the LBT spectra with a typical cadence of 220~s, we have selected a $\Delta t=440$~s. This gradient calculation is noisy, but it does avoid some of the pitfalls found in analyzing the residual time series.

The residual time-series (Figure~\ref{residuals}) analysis and the gradient time-series (Figure~\ref{instant}) method provide complementary information about the changes over the spectral time-series. The first displays the large-scale variability, while the latter provides information on changes localized in time and velocity. 

The broad features in the gradient time-series (Figure~\ref{instant}) appear in zig-zag patterns moving coherently in time and velocity. Unlike the accelerating features seen in some W-R star spectra, in \namex , some features appear to be decelerating and some can be followed across the line. Whether accelerating or decelerating, there is a fairly consistent slope to the gradient features of 7$\pm 2$~\kmss .

\begin{figure*}
    \centering
    \includegraphics[scale=0.35]{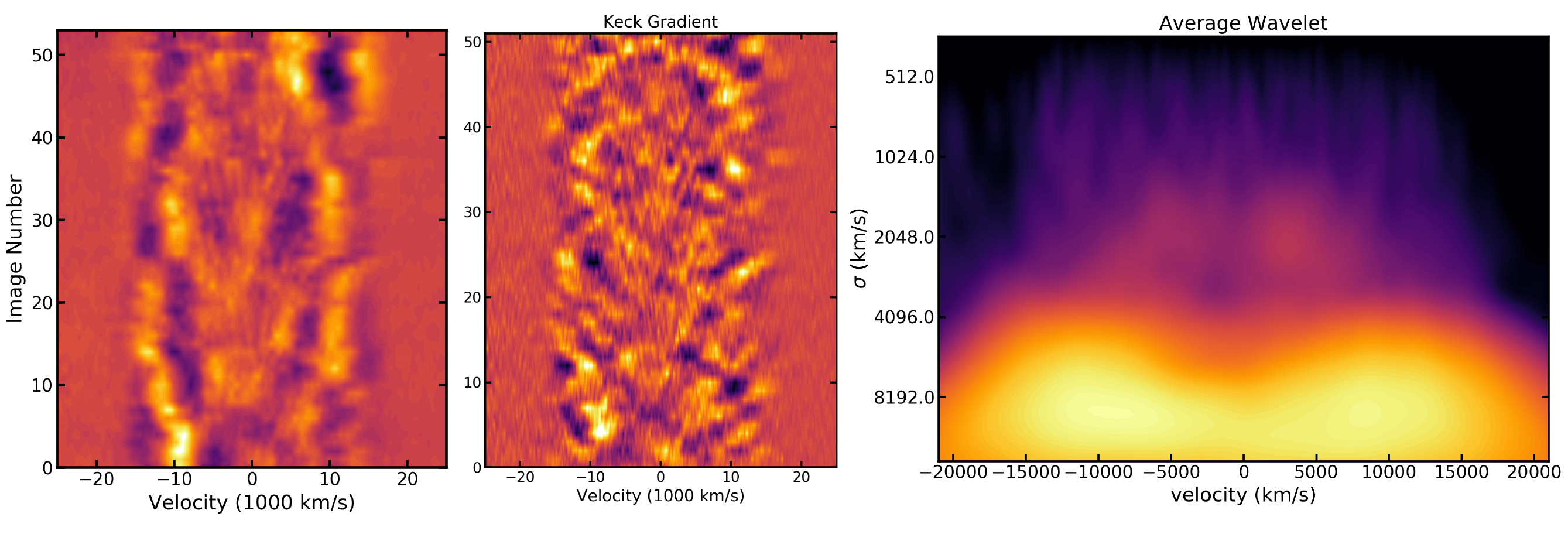}
    \caption{From left to right: the residual time series for the Keck observations, the instantaneous spectral variation gradient time series for the Keck observations, the scalogram of the wavelet transform of the Keck observations.}
    \label{keck_series}
\end{figure*}

Analysis of the longer Keck data set (Figure~\ref{keck_series}) confirms the features seen in the spectral time-series from the LBT data. The Keck gradient map clearly shows features that appear to move diagonally across O~VI line. The rate of 6$\pm 2$~\kmss\ is similar to that seen in the LBT data. Coherent features moving across the entire line is suggestive of corotating interaction regions (CIR) as seen in some W-R star winds \citep[e.g.][]{aldoretta16}. In \namex , the structures do not appear to survive for a full cycle. If this is an indication of rotation, the spin period would be approximately one to two hours. 

As observed in the LBT time-series, the Keck data also displays narrow features moving in velocity that are constrained to the central 5000~\kms\ of the O~VI line (Figure~\ref{keck_series}). 

\begin{figure*}[t]
    \centering
    \includegraphics[scale=1.1]{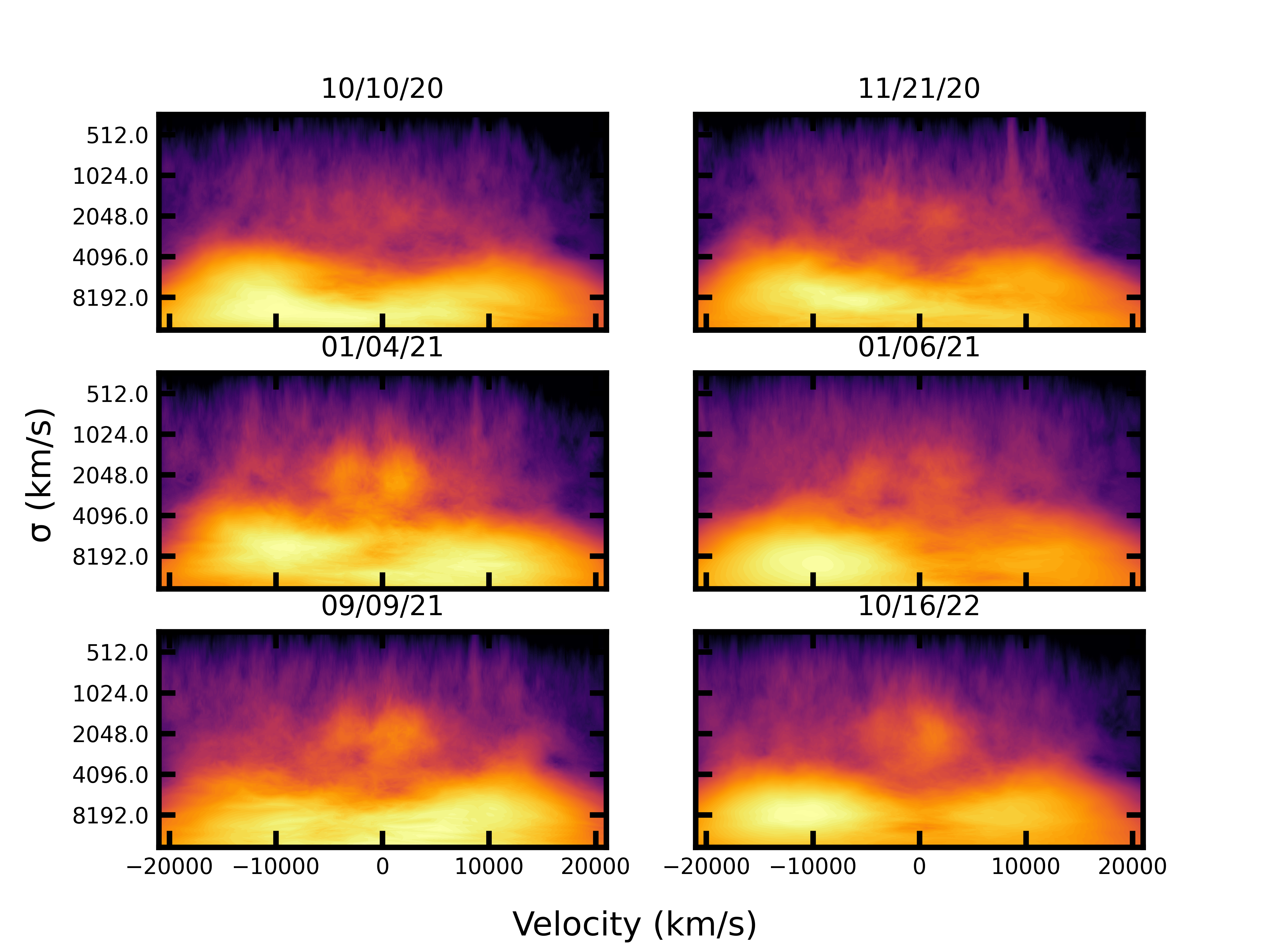}
    \caption{Scalogram plots of the wavelet transform of the residual time series for each LBT dataset. The x-axis displays the velocity relative to the O~VI rest wavelength of 3822{\AA}, and the y-axis shows the wavelet scale in \kms . The color represents the relative amplitude of the wavelet signal.}
    \label{scalograms}
\end{figure*}

\subsection{Wavelet Transform}

We applied a wavelet transform to the residual time-series in order to quantify the fluctuation scale across the emission line. The wavelet transform is implemented through PYTHON's {\it pywt} package \citep{wavelet19}. We found a high-order complex Gaussian wavelet (cgau7) provided the highest contrast scalograms. For a spectral time-series, the scalogram is calculated from each residual spectrum and these individual scalograms are averaged. Scalograms for the LBT data are shown in Figure~\ref{scalograms} and for the Keck time-series in Figure~\ref{keck_series}.

At the highest red-shifted and blue-shifted sides of the O~VI line, the scalograms quantify what was noticeable in the residual maps: the presence of broad fluctuations with size scales of 4000 to more than 8000~\kms . These broad fluctuations appear to weaken toward the center of the O~VI line. \footnote{Note that resolution of the scalogram decreases as the wavelet scale increases.  Thus, for fluctuations near the edge of the line, the scalogram power on the largest scales broadens beyond the limits seen in the data. The scalogram power that appears beyond $\pm 18000$~\kms\ is an artifact caused by the very wide wavelets when $\sigma> 8000$~\kms .}.

One striking feature of the scalograms is the isolation of the narrow features in $\sigma$ and line velocity. The residuals due to these narrow lines are generally contained within $\pm 5000$~\kms\ of the O~VI line center. The narrow features peak at wavelet scales of $\sigma\approx 2000$~\kms , that are consistent with the separation of the O~VI doublet. Further, in all the scalograms there appears a gap in the amplitude of the narrow features near zero expansion velocity of the O~VI line.

\begin{figure}[b]
    \centering
    \includegraphics[scale=0.40]{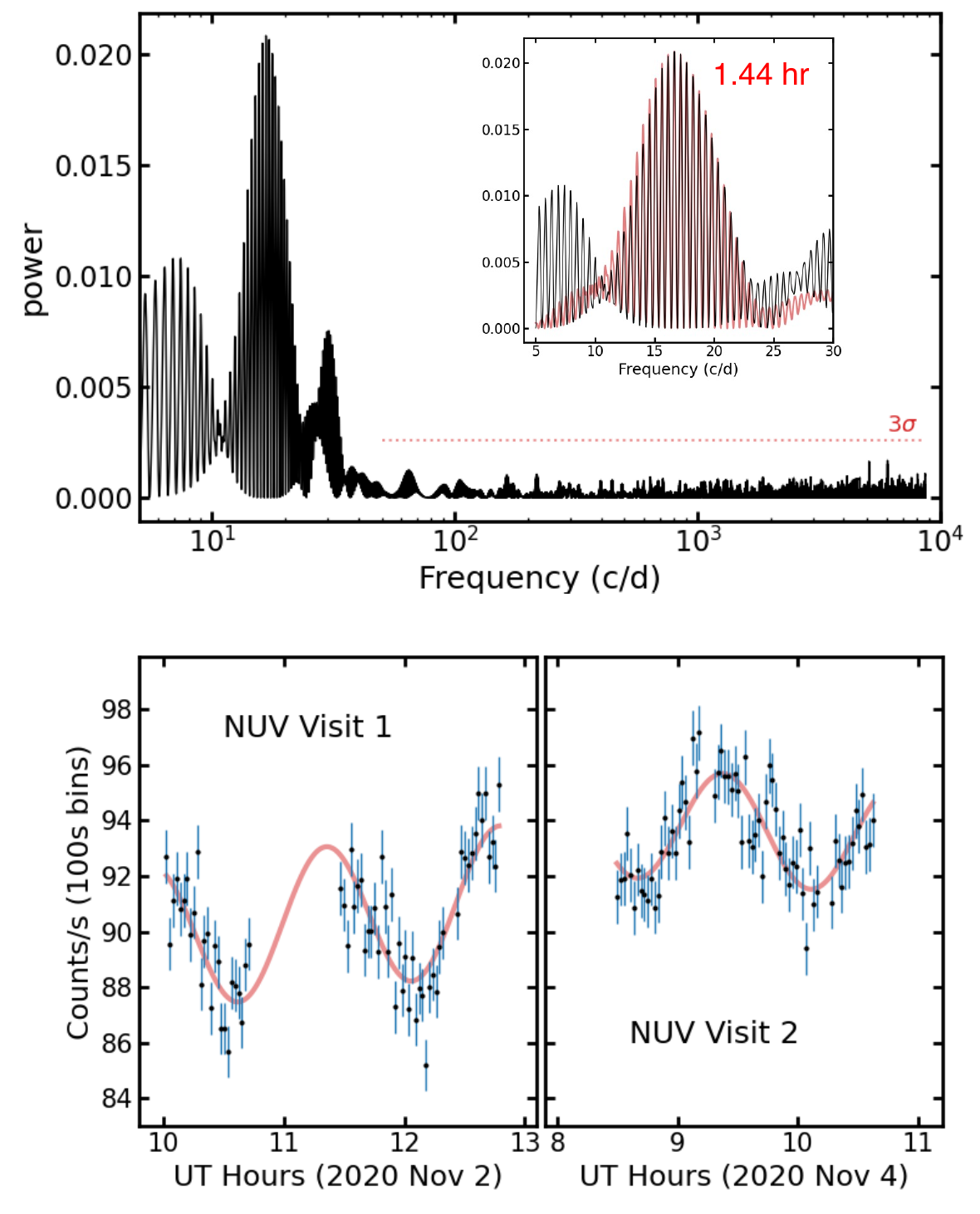}
    \caption{{\bf Top:} Periodogram of the full STIS NUV light curve constructed with 1~s bins. The dotted red line indicates the power required for a 3$\sigma$ detection of a periodic signal. The inset displays the frequency region zoomed-in on the power peak at 16.7~cycles~d$^{-1}$. Over-plotted in red is the window function resulting from the combination of the two HST visits separated by 48~hr.   {\bf Bottom:} The STIS NUV light curve averaged into 100~s bins. The 1.4~hr variability is clearly visible. A least-squares sinusoidal fit to each visit is displayed as solid red lines. The period of the fit sinusoid was fixed at 1.44~hr, but with the linear slope, amplitude and phase allowed to vary.     }
    \label{ls}
\end{figure}

\subsection{Photometric Variability} \label{sec:phot}

The recent merger of WDs is expected to produce a rapidly rotating remnant. Rotation periods between 10~seconds and 10~minutes are possible \citep{zhong23}. \citet{schaefer23} made an extensive analysis of photometric archives including TESS data with a 20~sec cadence, but no periodic photometric signal was identified. Instead, the optical variations suggest pseudo-random fluctuations in brightness. 

We downloaded the 120~s and 600~s cadence TESS Sector~52 light curves of \namex\ from the MAST archive. We also analyzed three stars within 10~arcminutes of \namex\ that also had 120~s cadence photometry available from Sector~52. We constructed the power spectral density (PSD) using the Lomb-Scargle method \citep[LS;][]{lomb76, scargle82}. The PSD is consistent with a $1/f$ or pink noise\footnote{The power spectrum of noise can be characterized by ${1}/{f^\alpha}$, where $f$ is frequency. The noise spectrum is referred to as `pink' for $\alpha\approx 1$, and `white' for $\alpha =0$.} \citep{pinknoise} where the PSD rises inversely with frequency out to 0.1 cycles~d$^{-1}$. The low-frequency limit is set by the length of the TESS sector and the high end by the Nyquist frequency. Caution is recommended in interpreting the low-frequency end of the TESS PSD as the simple aperture photometry (SAP) method shows jumps and long-term variability that is not intrinsic to the target.  Conditioned data (PDCSAP) reduces long-term variations, but can artificially add periodicities to the photometry. 

TESS sensitivity is weighted toward the long-wavelength end of the optical spectrum. In contrast, the STIS TIMETAG data allows analysis of variability in the NUV where a signal from the rotating remnant may be more clearly detected. We constructed a Lomb-Scargle periodogram from the background subtracted NUV count rate averaged in 1~s wide time bins which reach twenty times the Nyquist limit of the TESS data. The resulting periodogram is shown in Figure~\ref{ls}. There are no significant periodic signals detected at frequencies higher than 40~cycles~d$^{-1}$, corresponding to periods of less than 36~minutes.

There is a highly significant peak in the NUV periodogram at 16.68~cycles~d$^{-1}$. The STIS visits result in a complex window function, but the peak is consistent with a single periodicity of 1.44~hr. The light curve averaged into 100~s bins displays a clear oscillation with a peak-to-peak amplitude of 4\%\ (see the lower panel of Figure~\ref{ls}). While the single periodicity appears convincing, the small number of cycles over the observing windows allows for the actual variation to be quasi-periodic (QPO), or the unlucky sampling of pink noise.

\section{Discussion}

\subsection{Broad LPVs from the Wind}

\subsubsection{Possible Coherent Features in the Time-Series}

A rotating star with long-lived structures in its wind may generate spectral features that move across the full width of emission lines \citep{mullan84}. Co-rotating interaction regions (CIRs) have been detected in W-R stars winds \citep[e.g.][]{st-louis13}. For example, an extensive spectroscopic campaign studying WR~134 found a 2.25 day period in the He~II 5411~\AA\ line variations \citep{aldoretta16}. 
When the line variations are phased on this period, sinusoidal features are seen moving across the line, tracing the long-lived CIR similar to those predicted by \citet{dessart02}.

Coherent positive and negative features surviving for portions of the time-series are visible in the \namex\ gradient maps. This is most pronounced in the Keck gradient time-series shown in Figure~\ref{keck_series}. The time-series obtained with Keck has the longest continuous coverage and provides the best chance to follow long-lived features. One can follow positive and negative fluctuations that appear to move diagonally across the image. These apparently coherent features suggest a possible rotational period from the O~VI emission of roughly two hours.

\begin{figure}
    \centering
    \includegraphics[scale=0.47]{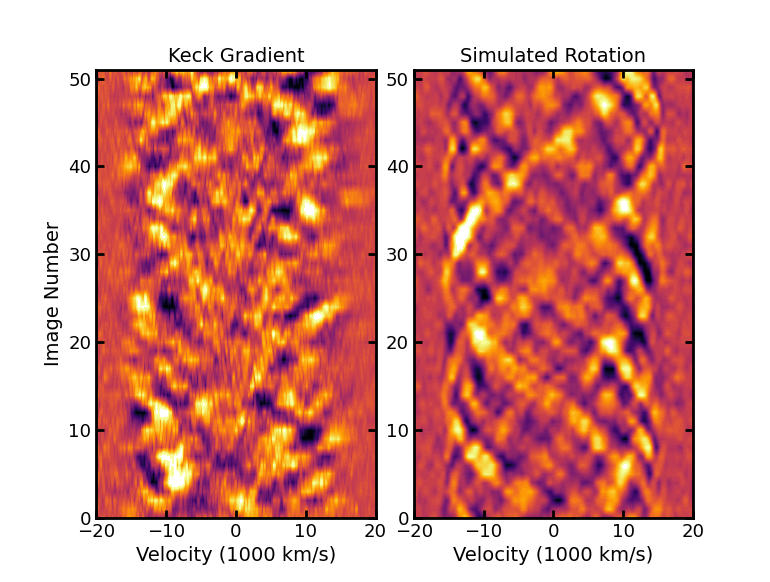}\hfill
    \caption{A comparison between the Keck gradient time-series and a gradient created from a simulated rotating shell. The shell has a random distribution of bright and dark spots that make one full rotation over the time-series. }
    \label{sim}
\end{figure}

\subsubsection{Toy Model of Rotation}

In an attempt estimate the rotation time-scale of the broad structures seen in \namex\ gradient maps, we created a simple, phenomenological simulation similar to the more detailed study by \citet{dessart02}. As discussed in Section~\ref{sec:line_shape}, the O~VI line can be approximated as spherical expanding shell of emission as the gas transitions from higher to lower ionization states and then fades at lower densities.  To simulate density variations within the shell, we add bright and dark fluctuations in the form of 2-dimensional circular Gaussian spots randomly placed on the surface of the shell. 

A simulated spectrum is generated by integrating the flux in rings centered on the z-axis and these fluxes are binned in radial velocity space. Once a raw spectrum is generated, it is convolved with a double Gaussian to match the broadening caused by the O~VI doublet.  To simulate the spectral time-series, the shell is rotated about an axis defined relative to the viewing direction, and the integration repeated to create subsequent spectra of the series.  We then follow the same data analysis processes as the actual observations by creating an instantaneous gradient time-series.

The Keck gradient time-series is compared to a simulation in Figure~\ref{sim}. Overall, the toy model of a rotating shell with fluctuating brightness patches is roughly congruent with the data and suggests an periodicity of about 2~hours.

We note that the rotational period of the stellar remnant is predicted to be tens of seconds to a few minutes \citep{zhong23}, close to the gravitational stability limit for a white dwarf. In comparison, a two-hour periodicity of CIRs appears very long, and this may not be directly related to the stellar rotation rate.

\subsection{Narrow Features in the O~VI Emission }

\begin{figure}
    \centering
    \includegraphics[scale=0.50]{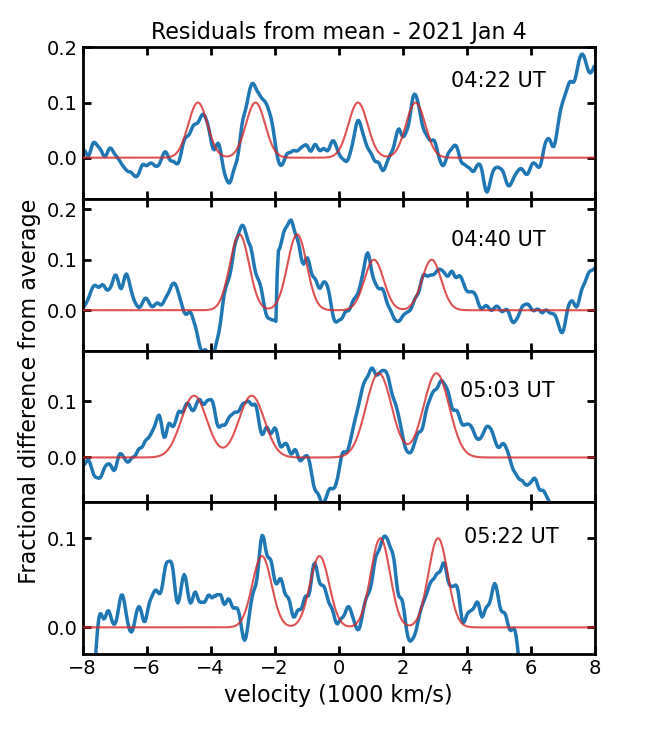}\hfill
    \caption{Four epochs from the LBT spectral series obtained 01/04/2021. Each panel is an average of three consecutive residual spectra and the time shown is UT of the middle spectrum. Solid blue lines are the data and the light red line shows simulated pairs of O~VI doublet emission features. Simulated doublets are Gaussian profiles with widths of 700~\kms\ (FWHM) and separated by a fixed 1800~\kms . The narrow doublet emission features appear at positive and negative velocities relative to the center of the broad emission line. }
    \label{narrow}
\end{figure}

\subsubsection{Origin of the Narrow Features in the Time-Series}

Narrow emission features seen in the time-series are a surprise and their presence suggests an additional component in the system with an origin distinct from the high-velocity wind. The narrow emission features come in pairs separated by about 2000~\kms , and this is suggestive of the O~VI doublet spitting of 1800~\kms. Figure~\ref{narrow} shows individual spectra where the pairs are clearly seen moving rapidly and unpredictably in velocity. Generally, there are two sets of emission pairs; one pair shifted to positive velocities and the other to negative.  

The individual lines have observed widths of $\approx 700$~\kms\ (FWHM), corresponding to velocity dispersions of 650 to 680~\kms\ after correcting for the spectrograph resolution.  The scalograms show that the narrow features are rarely present near zero velocity, as seen by a gap between the red-shifted and blue-shifted sides of the emission.

Properties of the O~VI 3811/3834~\AA\ doublet were calculated by \citet{sur07}. They found that the ratio of the transition probabilities is 1:1, suggesting that in emission the doublet components should be approximately equal in strength. Spectra of hot W-R stars and planetary nebulae show that the observed intensities of the two components are very similar \citep{ratio90,bond23}.

Resolving the O~VI doublet clearly establishes a low velocity dispersion to the emission, despite the velocity of the pairs shifted by up to 5000~\kms . Such a combination is difficult to explain in an expanding fast wind. However, a rotating disk or annulus seen at high inclination (nearly edge-on) naturally generates relatively narrow emission features separated by twice the disk rotational velocity. 

In Appendix~A, we consider that the narrow features could arise from Paschen-Back splitting from emission arising in a very strong magnetic field. We calculate that 5~MG fields can generate pairs of lines that would be found around $\pm2500$~\kms\ from the rest wavelength of the O~VI line. However, the observational properties of the narrow emission features are not a good match to the predictions of the Paschen-Back splitting.


\begin{figure}
    \centering
    \includegraphics[scale=0.50]{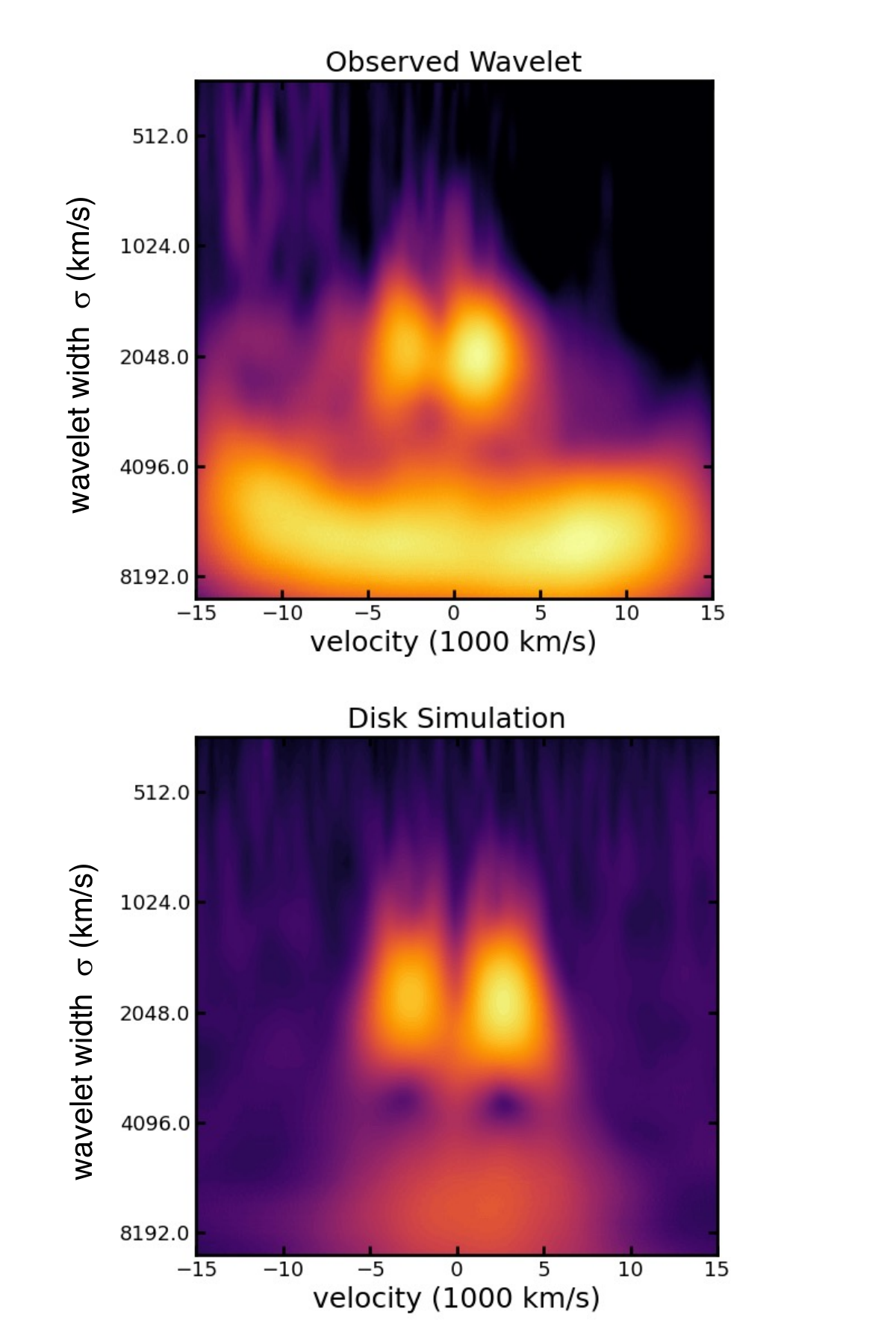}\hfill
    \caption{{\bf Top:} The wavelet scalogram of 60 instantaneous gradient spectra observed under good seeing conditions. The broad LPVs have been suppressed to better display the narrow features. This was accomplished by subtracting low-order spline functions fitted to each gradient spectrum before running the wavelet analysis. The observations show that the peaks of the narrow component are separated by 4200$\pm400$~\kms . The observations tend to be noisier on the blue-shifted side of the emission line due to the declining detector sensitivity towards short wavelengths. {\bf Bottom:} A simulated wavelet scalogram assuming a high inclination disk with a projected rotation velocity of $\pm 2500$~\kms\ and pseudo-random velocity fluctuations with a scale of 800~\kms .   }
    \label{simulation}
\end{figure}

\subsubsection{Narrow Spectral Features from a MHD Disk} \label{sec:disk}

The broad O~VI feature is consistent with a nearly spherical shell of emission, similar to what is seen in the winds of W-R stars. Instabilities in the wind generate the broad LPVs seen in the time-series spectra. 

We propose that the narrow emission features seen in the spectra arise from an unstable disk near the spinning, highly magnetic, stellar remnant. Figure~3 from \citet{ud-doula08} illustrates the geometry of the wind combined with the disk and equatorial outflow.

In contrast, the narrow features are constrained to velocities of less than $5000$~\kms\ and appear in pairs because O~VI is a close doublet (see Figure~\ref{narrow}). Emission from a disk or annulus can explain the low velocity dispersion that allows us to resolve the doublet, yet still generate high velocities on each side of the rest wavelength of $\pm 2000$ to $\pm 4000$~\kms . As discussed in the previous section, the extremely strong magnetic field on the WD merger leads to a large magnetic confinement parameter, $\eta_* >> 1$, and a Rigidly Rotating Magnetosphere (RRM) \citep{townsend05}.  

For an aligned rotator, a disk forms in the equatorial plane of the WD. Unlike a gravitationally supported accretion disk, the field and gas velocities increase with radial distance such that the observed projected velocity is
$v_p\; = \; R\, sin(i)\, 2\pi R_*/P_{wd}$, where $R=r/R_*$ is the scaled radial distance, $P_{wd}$ is the spin period of the WD, and $i$ is the inclination of the WD spin axis. For an inclination of 70$^\circ$ and a 1.0~M$_\odot$ WD spinning at half its breakup rate, $v_p\approx 2300R$~\kms . If the last closed field line is near $R\approx 2R_*$, then the highest rotational velocities would be at $\pm 4600$~\kms . Beyond the edge of RRM, the disk rotation velocities would decline as $1/R$ \citep{shenar14}, and retain velocities of $v_p> 2000$~\kms\ out to 4.5$R_*$. So an annulus of O~VI emission could be located well beyond the last closed field line, yet still have rotational velocities comparable to that at the WD equator.  To be visible in the scalograms, the disk velocities would need to be variable on a time-scales of several minutes. Given the breakout instabilities near the last closed field line, such a variability time-scale is plausible.

We simulated O~VI emission from a high inclination disk to see if the resulting scalograms were consistent with the observations. A projected disk velocity of $v_p=2500$~\kms\ was assumed and Gaussian line profiles were created pairs with the doublet separation of $1800$~\kms . Detection of the narrow features in the instantaneous gradient series requires velocity variability, so the lines from the disk were shifted pseudo-randomly with a scale of $\pm 800$~\kms .   The resulting simulated wavelet scalogram is shown in Figure~\ref{simulation}, and compared with the observed scalogram generated from combining two LBT nights with the best seeing.

Both the observed and simulated scalograms have peaks centered on wavelet scales of 2000$\pm300$~\kms , consistent with the separation of the O~VI doublet. In the observed scalogram, the peaks of the narrow component have a total separation of 4200$\pm400$~\kms , implying a disk rotation velocity of 2100~\kms\ at the location of the O~VI emission.

In the simulated scalogram, the red-shifted peak has a slightly higher amplitude than the blue-shifted one. This resulted from a difference in the randomly generated velocity fluctuations between the two sides. The instantaneous gradient detects changes in the velocity of the narrow features, so fluctuations with fewer velocity shifts lowers the wavelet amplitude. The peaks in the observed scalogram are also asymmetric, and this could be caused by similar random fluctuations in the motion of the narrow lines. More data is needed to test if this difference in amplitude between the red and blue-shifted peaks appears consistently in the narrow features.
The dip between the two peaks in the observed scalogram is slightly blue-shifted relative to the center of the broad O~VI emission. This may result from the two wavelet peaks having such different amplitudes.

Overall, a turbulent, high-inclination disk with a rotation velocity of 2000~\kms\ to 3000~\kms\ at the location of the O~VI emission could account for the narrow features detected in \namex .


\subsubsection{Breakout Instabilities in Magnetohydrodynamic Models}

Magnetohydrodynamic (MHD) modeling in two-dimensions (2-D) has been extensively applied to simulate magnetically channeled winds from massive stars where the initial dipole is aligned with the stellar rotation axis \citep[e.g.][]{ud-doula08}. It has recently become possible to make three-dimensional MHD calculations of magnetically channeled winds from non-aligned magnetic rotators \citep{subramanian22}. \citet{zhong23} calculated 2-D MHD wind models using the parameters expected for the \namex\ merger candidate. However, stellar parameters such as mass, rotation period, and magnetic moment have not been well constrained observationally.

In general, an aligned rotator generates an approximately spherical wind combined with a time-varying disk structure. The field lines are closed within the co-rotation radius, and plasma flowing from the opposite magnetic hemispheres results in a density enhancement at the equator. Some of this gas falls back toward the star and some is held near the last closed magnetic loop.  The build-up of density eventually drags the field out into an expanding disk \citep{ud-doula08} in a breakout instability. Magnetic reconnection also plays a part in the breakout as the field lines above and below the disk have opposite polarities. The breakout instability has been used to explain X-ray variability in massive stellar winds \citep{ud-doula06}, but it may be a source of the optical/UV variations observed in \namex\ \citep{zhong23}.

Guided by MHD simulations, \citet{ud-doula08} made analytic estimates of the disk breakout time-scales for magnetic, rotating OB stars based on the work of \citet{townsend05}. Equation~19 in \citet{ud-doula08} shows that the breakout time-scale, $t_b$, depends on the free-fall time, $t_{ff}$, the magnetic confinement parameter at the stellar surface, $\eta_*$, and an estimate of the radial distance of the last closed magnetic field line in the disk.

For an estimate of the disk breakout time-scale, we assume the stellar radius of $R_*= 0.008R_\odot$ consistent with a WD mass of 1.0$~M_\odot$ \citep{nauenberg72}. The mass of \namex\ is uncertain, but it is roughly around a Solar mass. We also assume the WD is spinning at half its breakup rate\footnote{Defined in \citet{ud-doula08} as $W=0.5$, where $W$ is the rotation parameter defined as the equatorial velocity divided by the orbital velocity at the stellar surface.} as this keeps the star close to spherical symmetry. Inserting these values into the \citet{ud-doula08} parameterization, the disk breakout time-scale is:
\begin{equation}
t_b\; \approx\; \frac{3.6\; \eta_*}{\xi_0(\xi_0^3-1)} \;\; (s),
\end{equation}
where $\xi$ is the radial distance scaled by the co-rotation radius, $R_K$, the radius where the velocity of the magnetic field matches the Keplerian orbital velocity. $\xi_0$ corresponds to the last closed magnetic field line in Kepler radii. From the MHD calculations by \citet{zhong23} (their Figure~1), the last closed field line reaches roughly to 2.0~$R_*$, corresponding to $\xi_0 = 1.3$, and reducing Equation~2 to $t_b\approx 3\eta_*$ in seconds.

The magnetic confinement parameter, $\eta_*$, is the ratio of the magnetic energy density and the wind energy at the stellar surface, and depends on the square of the poorly constrained surface magnetic field \citep{ud-doula06}. After merging, a remnant WD is predicted to  have a field strength as high as 10$^8$ to 10$^9$~G \citep{beloborodov14,gvaramadze19}. For the parameter values estimated in \citet{kashiyama19}, that is, a surface field of $2\times 10^7$~G and a mass-loss rate of $3\times 10^{-6}$~M$_\odot$yr$^{-1}$, the disk breakout time-scale is roughly 2000~s. This is comparable to the 5000~s pseudo-periodic variation observed in the HST UV data (Section~\ref{sec:phot}).

Relatively small changes in the assumed WD magnetic field results in a wide range of breakout time-scales. For example, applying the parameters of the \citet{zhong23} fiducial model, B1.5e6$\Omega$0.23, results in $t_b=25$~s. This value is very close to the instability time-scale found in their MHD simulation, and coincidently, the WD spin period in their model. 

\citet{ud-doula08} suggests that a "hierarchy" of time-scales may result from breakouts occurring at varying radii in the disk. In an active disk, the location of the last close loop could be highly variable, and the breakout time-scale declines steeply with radius (see Equation~2). Such a range of time-scales could be the source of the pink noise apparent in the optical light curves.

\section{Summary}

\begin{itemize}

\item We present near-IR spectra of \namex . Clear emission features are seen in the $H$ and $K$ spectral bands (Figure~\ref{irspec}), and we associate the sharp features detected at each end of the $K$ band with Ne~VIII lines.

\item We analyzed the O~VI line profile and find that its structure is consistent with an optically thin spherical shell of emission. O~VI emission begins at a velocity of $v_{min}=11400\pm 200$~\kms\ and the wind has a terminal velocity of $v_\infty = 17000\pm200$~\kms . The results for the O~V emission line in the UV are similar except that the emission begins further out at $v_{min}=12900\pm 300$~\kms , consistent with acceleration and cooling. 

\item Time-resolved spectra of the O~VI emission shows broad line profile variations with typical amplitudes of $\pm10$\%\ of the line flux. The LPVs in the O~V NUV line have amplitudes of $\pm15$\%. The widths of the broad LPVs are roughly 6000~\kms , and these are wider than seen in typical W-R star winds which tend to be 5\%\ to 10\%\ of the full line width \citep{lepine99a}. 

\item The gradient time-series spectra suggest both acceleration and decelerating features with slopes of about 7~\kms~/s. Some LPVs appear to traverse the line and then even reverse direction, although they do not remain coherent for an entire cycle. These may be co-rotating interaction regions associated with the wind \citep[e.g.][]{dessart02}. Extrapolating from the rate these LPVs traverse the O~VI line suggests a rotation period at the O~VI shell of a few hours. 

\item The O~VI emission also displays narrow features that are distinct from the broad LPVs. These features are sufficiently narrow that the O~VI doublet is resolved, so they appear as emission pairs separated by $\approx1800$~\kms . Accounting for the spectroscopic resolution, the features have a width of only 650~\kms . The doublets rapidly change velocity, but the scalograms show that they are constrained to the inner $\pm 5000$~\kms\ of the broad O~VI line. We propose that these narrow features originate in a disk generated by a RRM. The observed narrow O~VI emission requires disk rotation velocities of order $2000/sin(i)$~\kms .
This rotation rate naturally arises from a WD spinning near breakup, and the RRM would maintain those high rotation velocities out to several WD radii.  

\item As shown by \citet{schaefer23}, optical variability in \namex\ does not show any consistent periodicity on time-scale longer than 40~s, the Nyquist limit of the TESS photometry. Instead, optical variability appears consistent with pink noise where the amplitude of random fluctuations increases with decreasing frequency. Analysis of HST photometry in the NUV shows no significant periodicity on time-scales between 2~s and 1800~s. A strong 1.4~hr periodic or pseudo-periodic signal is detected in the NUV. However, given the limited time coverage of the photometry, this may be unlucky sampling of the pink noise. Further UV photometry is needed.

\item We investigated the possible signature of a magnetic field in the PB regime to explain the narrow features observed in the O~VI emission. While velocities of order $\pm4000$~\kms\ are achievable from the $\sigma$ components for fields of 6~MG, their pair separations appears too small to explain the $2000$~\kms\ splitting observed in the narrow features. The narrow features are better matched by the unperturbed O~VI doublet separation of 1800~\kms . 

\end{itemize}

\namex\ is a dynamic star with a fast wind that display strong instabilities. The broad LPVs hint of coherent structures in the wind that may signal rotation several stellar radii from the WD. Further, we detect a narrow component to the O~VI emission line and propose that it originates from a disk outflow generated in a rigidly rotating magnetosphere of the rapidly rotating WD. 

\begin{acknowledgements}

We thank D. Balsara and C. Littlefield for helpful discussions.

The LBT is an international collaboration among institutions in the United States, Italy and Germany. LBT Corporation partners are: The University of Arizona on behalf of the Arizona Board of Regents; Istituto Nazionale di Astrofisica, Italy; LBT Beteiligungsgesellschaft, Germany, representing the Max-Planck Society, The Leibniz Institute for Astrophysics Potsdam, and Heidelberg University; The Ohio State University, representing OSU, University of Notre Dame, University of Minnesota, and University of Virginia

\end{acknowledgements}

\facilities{LBT, HST (STIS), TESS}

All the {\it TESS} data used in this paper can be found in MAST: \dataset[10.17909/q1nt-xv95]{http://dx.doi.org/10.17909/q1nt-xv95}

All the {\it HST/STIS} data used in this paper can be found in MAST: \dataset[10.17909/dp4s-na11]{http://dx.doi.org/10.17909/dp4s-na11}

\software{IRAF \citep{tody86,tody93}, pyraf \citep{pyraf12}, pywavelet \citep{wavelet19}}

\appendix

\section{Narrow Features from a Strong Magnetic Field}

The magnetic field at the WD surface could be quite large,
with estimates as high as 10$^8$~G \citep[e.g.][]{gvaramadze19}, so the field strength at the disk breakout radius around $2R_*$ might reach 10~MG. The Zeeman effect may impact the narrow O~VI emission if it is originating near the star.

\citet{lykou23} constrained the magnetic field in the wind based how Zeeman splitting could contribute to the observed line widths. They found the field must be less than 2.5~MG at the distance of the O~VIII 606.4~nm line. They note that the field strength falls of quickly with radius, so that the field at the stellar surface could be significantly higher than this limit.

With three bound electrons, the O~VI ion is a lithium-like atom. The 381.1/383.4~nm  doublet comes from fine-structure splitting of the $^2P$ levels with the transition down to the $^2S$ level. This set of transitions has the same quantum numbers as the 670.7~nm doublet in Li~I, where the $^2P_{\frac{1}{2}}$ and $^2P_{\frac{3}{2}}$ levels differ by $\Delta E=$0.34~cm$^{-1}$ \citep{kochukhiv08,stift08}. At modest field strengths, the anomalous Zeeman effect divides the doublet into ten components. The Paschen-Back (PB) regime dominates at very high magnetic field strengths where the structure evolves to essentially five emission lines. The transition between the anomalous Zeeman and PB regimes occurs for fields reaching $B_{pb}\approx \frac{\Delta E}{3\mu_B}$, where $\mu_B$ is the Bohr magneton. For the 670.7~nm Li~I line, this transition begins around 3~kG \citep{kochukhiv08} 

\begin{figure}
    \centering
    \includegraphics[scale=0.60]{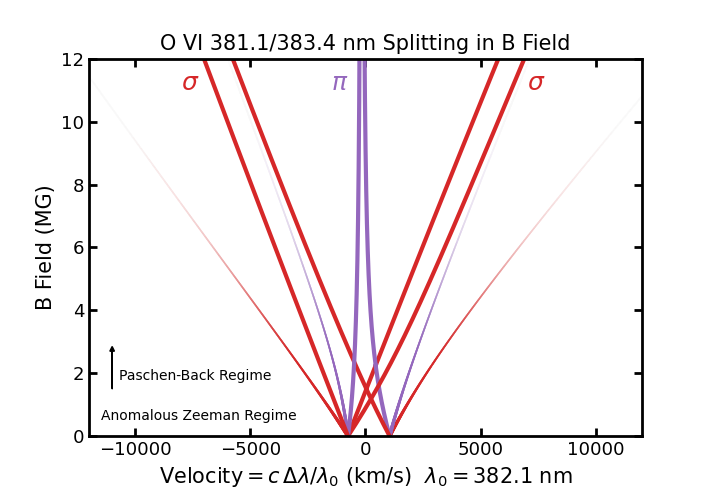}\hfill
    \caption{The Zeeman/Pashen-Back line splitting versus magnetic field for the O~VI doublet. The wavelength changes have been converted to apparent velocity shifts assuming a ``rest'' wavelength of 382.1~nm between the unperturbed doublet lines. Four of the ten anomalous Zeeman lines fade at high field strengths leaving three pairs of Paschen-Back lines.  For B~fields around 5~MG, the pairs of $\sigma$ components reach apparent velocities of about 2500~\kms , similar to what is seen in the narrow O~VI features. }
    \label{zeeman}
\end{figure}

While the physics of the Li~I and O~VI doublets is comparable, the fine structure splitting of the doublet is much larger in O~VI. With a $\Delta E=156$~cm$^{-1}$, the onset of the PB regime is about 1~MG for the 381.1/383.4~nm doublet. We calculated the $^2S$-$^2P$ energy level eigenvalues for the Li~I and O~VI doublets as a function of magnetic field strength \citep{landi04,sowmya14} and plot the results for O~VI in Figure~\ref{zeeman}. Our calculation of the PB effect for Li~I matched the results published by \citet{stift08}, providing confidence that the O~VI shifts are valid. The PB regime results in
three pairs of lines, with the closely spaced $\pi$ components remaining near the doublet's central wavelength. The $\sigma$ component pairs are separated by velocities of 1000~\kms\ to 1400~\kms , somewhat smaller than that of the original fine structure splitting of 1800~\kms .  

The PB regime of the O~VI emission is intriguing as fields of 6~MG appear able to generate pairs of lines separated by $\gtrapprox 
1000$~\kms\ at velocities that appear shifted from rest by $\approx\pm 4000$~\kms . These properties are similar to the narrow lines observed in \namex . Modest variations in the field strength would move the centers of the $\sigma$ component pairs in velocity, but leave the their spacing nearly constant. The $\sigma$ components shift by 500~\kms\ when the field is varied by 1~MG in the PB regime. The $\pi$ component would not vary with changes in the field and likely be suppressed in our gradient time-series that is designed to detect variability.  However, the strong fields required for the PB regime would likely arise within a few radii of the WD, where high velocities would Doppler broaden and blend the emission line pairs. 

The $\sigma$ component line pairs calculated from the PB regime tend to have smaller separations than the unperturbed doublet. The fine structure splitting of the O~VI lines without a magnetic field is 1800~\kms , while the $\sigma$ pairs are split by $1000$ and $1400$~\kms . The wavelet analysis of the time-series spectra show peaks at $\sigma \approx 2000\pm 300$~\kms , which is consistent with the doublet separation, but marginally inconsistent with separations in the PB regime. While explaining the observed narrow features with a strong field in the PB regime is intriguing, the data currently does not support this idea.

Clearly resolving the O~VI doublet implies that the region producing the narrow emission likely has a magnetic field strength of less than $1$~MG so that the anomalous Zeeman splitting does not wash out the peaks.



\end{document}